# Personalization in E-Grocery: Top-N versus Top-k Rankings

Franziska Scherpinski, Stefan Lessmann [1]

*Humboldt-University of Berlin, Unter den Linden 6, 10099 Berlin*

**Abstract:** Business success in e-commerce depends on customer perceived value. A customer with high perceived value buys, returns, and recommends items. The perceived value is at risk whenever the information load harms users' shopping experience. In e-grocery, shoppers face an overwhelming number of items, the majority of which is irrelevant for the shopper. Recommender systems (RS) enable businesses to master information overload (IO) by providing users with an item ranking by relevance. Prior work proposes RS with short personalized rankings (top-k). Given large order sizes and high user heterogeneity in e-grocery, top-k RS are insufficient to diminish IO in this domain. To fill this gap and raise business performance, this paper introduces an RS with a personalized long ranking (top-N). Undertaking a randomized field experiment, the paper establishes the merit of shifting from top-k to top-N rankings. Specifically, the proposed RS reduces IO by 29.4% and lowers users' search time by 3.3 seconds per item. The field experiment also reveals a 7% uplift in revenue due to the top-N ranking. Substantial benefits for the customer and the company highlight the business value of top-N rankings as a new design requirement for recommender systems in e-grocery.

## 1. Introduction

Business Analytics comprises information technology and quantitative empirical models to support processes and decision-making in business organizations (Delen and Ram 2018). Aiming at improving efficiency and decision quality, the overarching goal of business analytics is to raise business performance (Delen 2015). Business performance relies on customer perceived value (CPV). Customers who perceive a service/product as valuable display higher satisfaction (Abu ELSamen 2015; Chang and Wang 2011; Kuo et al. 2009; Yang and Peterson 2004), purchase intention (Chang and Wang 2011; Chen and Dubinsky 2003; Kuo et al. 2009; Lien et al. 2015), and loyalty (Abu ELSamen 2015; Chang and Wang 2011; Yang and Peterson 2004). A threat to the CPV is information overload (IO). IO occurs when the amount of information exceeds the customer´s cognitive processing capacity (Milord and Perry 1977). A customer requires more processing capacity the higher the cognitive load is, e.g. the number of viewed items. In e-commerce, IO often occurs to a user who is scanning a commercial web page with thousands of items. IO yields negative effects on user´s shopping behavior, i.e. users are confused by too much information (Jacoby et al. 1974b). The negative effects of IO on users reduce the CPV and thereby harms business performance. A decrease in the information load reduces these negative effects (Eppler and Mengis 2004). Therefore, businesses aim at reducing IO to increase CPV and business value.

---

[1] Corresponding authors: {franziska.scherpinski, stefan.lessmann} @hu-berlin.de

This chain of effects makes IO reduction relevant to many e-commerce businesses. IO occurs in particular in online grocery, since a user faces thousands of items of which 99% are irrelevant (Levitin 2015). The number of items in a grocery store has increased massively in the past decades from around 8.000 items in the 1970s (Jacoby et al. 1974a) to more than 40.000 items nowadays (Yuan et al. 2016). The electronic commerce of grocery items (e-grocery) is a fast-growing industry, in particular in the United States (PitchBook 2017) with global players like Amazon Fresh and Walmart Online Grocery. E-grocery can address IO with business analytics while offering full assortment. The online environment and rich customer data enable the integral application of Recommender Systems (RS). RS are a viable approach to address IO (Aljukhadar et al. 2010; Jannach et al. 2012; Schafer et al. 1999) and play a significant role as user-specific decision-making systems in business analytics (Patnaik and Hiriyannaiah 2017). The persistence of IO in e-grocery and the increasing practical relevance motivate our research on e-grocery RS.

This paper has the objective to reduce IO in e-grocery by designing and field testing a top-N RS. We identify large orders and user heterogeneity as primary domain-specific problems that aggravate IO. Large orders demand the user to search for many items in an online shop. High user heterogeneity leads to large grocery assortments, whereas the vast majority is irrelevant to an individual user. The number of irrelevant items the user faces in each of many search processes results in an unnecessary high cognitive load, thus to IO.

Based on the requirement engineering, we deduce the RS design requirements of collaborative filtering and a personalized, long ranking (top-N) to minimize the cognitive load. This requirement allows all items the user aims to buy are in a personalized ranking. The state-of-the-art RS in e-grocery are collaborative and personalized but have short rankings (top-k). A user can find only a fraction of relevant items in a top-k ranking and must browse a non-personalized or semi-personalized ranking for the remaining items the user wants to purchase. Less personalized rankings increase the number of irrelevant items in a user's shopping journey. Thus, a user suffers from more cognitive load during the search processes to find all items. A practical example from Walmart Online Grocery illustrates the weakness of top-k rankings. Yuan et al. (2016) use a personalized top-8 ranking with 12% accuracy. This means each user can expect to find ~1 item within the top-8 ranking. Though, customers at Walmart Online Grocery usually buy 21 and up to 70 items (Yuan et al. 2016). This means users need to navigate through a non- or semi-personalized ranking with even less than 12% accuracy to buy the remaining up to ~69 items, which aggravates IO.

To evaluate the effect of the new top-N RS requirement, we represent the number of ranked items by the two extremes k and N with $k \ll N$. Top-k RS denotes an RS that ranks a small number of items (e.g.,7) and top-N RS represents a large number of items (e.g.,4000) respectively. This understanding of top-k and top-N RS facilitates quantifying their respective potential to reduce IO later in the paper. Since an e-grocer's assortment is very large, both rankings must be complemented by a non-personalized or semi-personalized ranking for the remaining items in the grocery store.

The contributions of the paper intend to evaluate the potential of top-N RS in e-grocery and motivate their application in practice. Thus, beneficiaries of the paper are RS researchers and e-grocery practitioners who intend deploying building RS. Specifically, the paper contributes to the literature by identifying a new design requirement for e-grocery RS (i.e., top-N ranking) and providing the corresponding methodology. Further, we contribute original empirical results from evaluating the effects of top-N RS in comparison to top-k RS in e-grocery with an A/B test. The field experiment detects a significant IO reduction of 29.4% and additional positive effects on users and companies like a search time reduction by 3.3 seconds and revenue increase by 7%.

## 2. Methodology

In this section, we define the application environment of our study and explain how we implement the design science research paradigm.

The application environment is a ranking problem by user preferences for the full e-grocery assortment. A ranking problem exists if the RS aims to rank the items of a merchant (Aggarwal 2016). Since we aim to reduce IO, we need to facilitate the user's usual grocery selection process. Thus, we focus on user preferences. This focus excludes food RS consulting on recipes, meal plans, menus, or items based on health aspects like nutritional needs. Ranking the full assortment of an e-grocer means that the RS not only recommends an assortment subset, like new items (Lawrence et al. 2001). Additionally, we concentrate on RS using data that every e-grocer stores to provide smooth and broad applicability. For this reason, we exclude RS asking the user to explicitly state preferences, like ratings, interests, or search terms.

Design science focuses on creating and evaluating innovative IT artifacts, like algorithms, that enable organizations to address critical information-related tasks (Hevner and Chatterjee 2010). We create and evaluate a top-N RS that enables e-grocers to address IO through traversing the design cycle (build & evaluate), relevance cycle (requirements & field testing), and rigor cycle (groundings & additions to the knowledge base). First, we review the existing IO and RS literature (grounding). After formally defining the problem, the requirement engineering section uses domain characteristics to identify design requirements for an RS in e-grocery (requirements). Subsequently, we map existing algorithms against deduced requirements to identify the research gap. To fill in the gap, we adapt an RS from another domain to e-grocery that meets all derived requirements (build). An offline evaluation validates that the adaptions meet the requirements (evaluate). Afterward, we apply a top-k RS and the developed top-N RS online at an e-grocer to evaluate effects on IO, users, and the company (field testing). Eventually, all effects of the top-N RS in the field are summarized and discussed (additions to the knowledge base). Figure 1 rests on the findings of Hevner (2007) and displays in which order we traversed the design science research cycles.

The majority of food RS research provides only offline evidence (Trattner and Elsweiler 2017). These offline evaluations do not reflect real-user behavior and thereby risk valid assessment. One significant advantage of our study is that we not only perform an offline evaluation but also run a large scale randomized online experiment in the field. Figure 1 connects the design science research evaluation steps to the applied evaluation methods. By realizing an online evaluation, we aim at a comprehensive assessment to overcome existing research weaknesses.

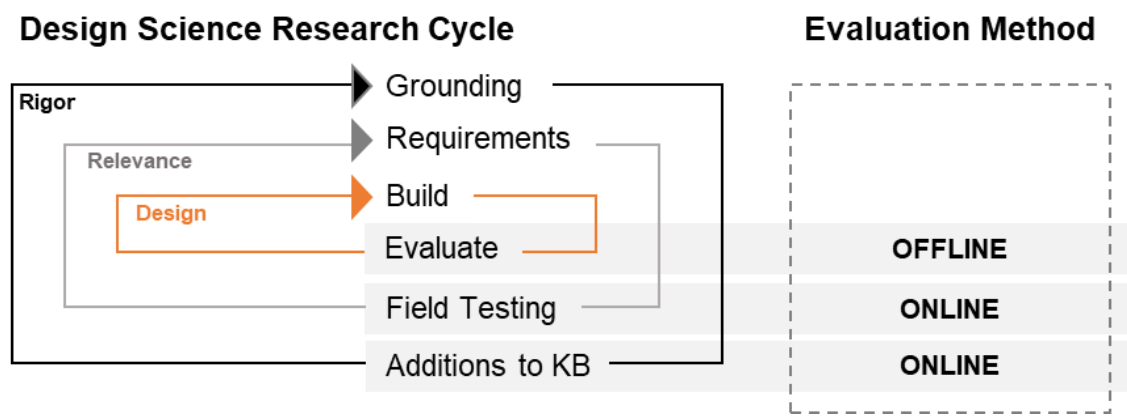

Figure 1. Design science research cycles and evaluation method

In summary, we ground our research on the application environment of solving a ranking problem by user preference for the full e-grocery assortment. For this reason, we focus on testing a new top-N RS in the field and evaluate its effect on IO, users, and the company.

## 3. Literature Review

IO has a long history in various fields of research (Eppler and Mengis 2004). Various overlapping definitions of IO exist (Melinat et al. 2014). RS address the field of consumer research. Authors in consumer research define IO as the condition when information supply exceeds the processing capacity of an individual (Eppler and Mengis 2004; Jacoby et al. 1974b). Applying the concept of IO to e-grocery, the amount of information is proportional to the number of items the user processes to complete the shopping journey. The processing of an item is the comparison with the items on a shopping list. This comparison enables the user's purchase decision. Therefore, the amount of information corresponds to the number of (purchase) decisions a user takes. Thus, IO occurs when the amount of decisions exceeds the processing capacity of an e-grocery user. A reduction of information load reduces the harmful effects of IO. Below, we focus on RS literature that addresses IO in e-grocery.

Prior work has designed RS for the defined application environment. These RS use methods from all three major fields of RS research: collaborative (Dias et al. 2008; Li et al. 2009; Sano et al. 2015; Wan et al. 2018; Wu and Teng 2011), content-based (El-Dosuky et al. 2012), and hybrid-based filtering (Yuan et al. 2016). All of them provide personalized recommendations and apply diverse types of evaluation. El-Dosuky et al. (2012) perform a user study with five users. The majority of authors evaluate their RS offline (Li et al. 2009; Ma et al. 2019; Sano et al. 2015; Wan et al. 2018) and some authors online (Dias et al. 2008; Yuan et al. 2016). These papers use classification metrics like hit rate (Li et al. 2009; Yuan et al. 2016) as well as precision, recall, and F-measure (El-Dosuky et al. 2012; Sano et al. 2015; Wan et al. 2018), ranking quality measures like nDCG and mAP (Ma et al. 2019) or do not provide information (Dias et al. 2008). Some authors (Panikar et al. 2016; Wu and Teng 2011) only suggest and exemplify their algorithm without any evaluation. In RS research, the term top-k refers to the overall length of a ranking (Aggarwal 2016). In the case of a top-3 ranking, the RS recommends the three items ranked first. The e-grocery RS in the literature range from top-3 (Dias et al. 2008; Li et al. 2009; Wu and Teng 2011) to top-32 (Wan et al. 2018). To our knowledge, these nine top-k RS build the state-of-art in the defined application environment. Since there is no top-N RS for e-grocery in literature, there exists no comparison between top-k and top-N RS up to now.

In summary, IO occurs when the amount of decisions exceeds the processing capacity of an e-grocery user. There are nine state-of-the-art top-k RS in the defined application environment that may be able to address IO.

## 4. Problem Statement

In the following, we describe the presence and adverse effects of IO in e-grocery. Sequentially, we illustrate user heterogeneity and large orders as two root causes of IO identified in e-grocery. User heterogeneity increases the amount of information in terms of assortment size at an e-grocer. Large orders require the user to process a proportional volume of information.

In e-grocery, users have to choose from several thousands of items to fill their baskets with 21 unique items on average (Yuan et al. 2016). The large item-to-user-ratio, combined with users' limited processing capacity, confirms IO in grocery shopping (Jacoby et al. 1974b; Levitin 2015). IO has adverse effects on e-grocery users. Eppler and Mengis (2004) cluster results of several IO studies to four main symptoms or effects: suboptimal decisions; strenuous personal situations like demotivation, stress, and confusion; limited information search and retrieval strategies; arbitrary information analysis and organization.

First, IO drivers are high inter- and inner-user-heterogeneity. The inter-user heterogeneity originates from the massive potential user base. Anyone requires grocery items in their daily life. Millions of people have access to the internet and could be part of the e-grocery user base. Theoretically, e-grocery should offer all required basket configurations for this heterogeneous user base. To exemplify complexity, a study in the US found that out of 32 million shoppers, there has been no pair of shoppers that bought the same array of items over a year (Catalina 2013). This individuality of shopping baskets highlights the user heterogeneity. The inner-user heterogeneity originates from the fact that grocery shopping is multitasking (I purchase food while buying laundry detergents) and multi-people (I buy groceries for the entire family) activity (Yuan et al. 2016). Inner-user heterogeneity leads to a broad set of items demanded by a user. This inter- and inner-user heterogeneity requires a wide assortment and results in a high amount of information the user must process. Consequently, user heterogeneity is one cause of IO in e-grocery.

The second e-grocery cause of IO is the characteristic of large orders (Yuan et al. 2016). An order consists of all items on a physical or theoretical shopping list. In the following, we reason that the length of the shopping list s determines the number of required purchase decisions. For simplicity, we also assume that a user can find all items on the shopping list in one store and then stops the shopping journey. Thus, the shopping list and final order consist of the same items and length. Figure 2 displays 10 items on a grocery shelf (middle row) as an example. The upper and lower row represents two shopping lists differing only in the number of items. Assume, a user can pass the shelf from left to right or the other way around. In each scenario, a user compares an item from the shelf to each item on the shopping list (shown by a line). The comparison enables the user to make purchase decisions for this item.

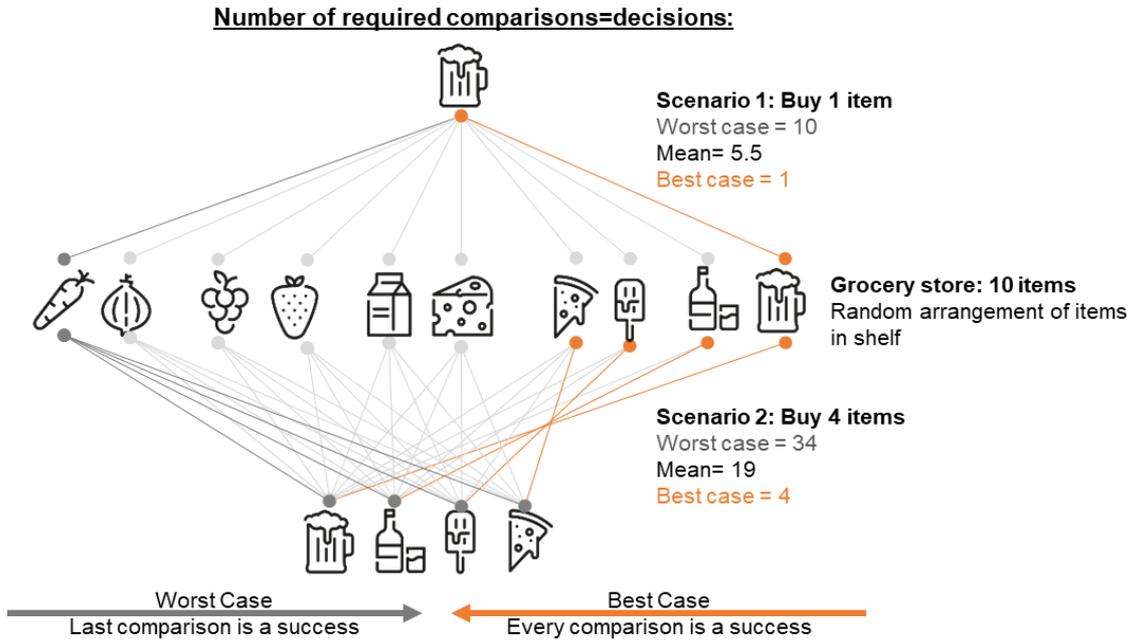

Figure 2. Large orders aggravate information overload in e-grocery

When a user passes the shelf from right to left (best-case ranking), every decision is a hit for both shopping lists. Accordingly, the number of required decisions is minimal, i.e., 1 and 4. The best-case ranking demonstrates that the order size increases the number of decisions in any case. The number of decisions is maximal, in case the user passes the shelf from left to right (worst-case ranking). The number of required decisions in the worst and best case ranking denoted as $D_{bc}$ and $D_{wc}$, is calculated via equation (1) and (2) with a as the size of assortment and s as the size of the shopping list/order.

$$D_{bc} = \sum_{i=0}^{s} s - i \quad (1)$$

$$D_{wc} = (a-s)*s + \sum_{i=0}^{s} s-i \quad (2)$$

For all equations, the number of items put on the shopping lists drives the number of decisions. This example clarifies that large order sizes – the typical scenario in e-grocery (Yuan et al. 2016) – result in high information load for a user. Of course, items are usually clustered in categories at a grocery store, like beverages and frozen goods, and not randomly distributed. This decreases the number of comparisons since the user first compares items from the shopping list to product categories. However, the amount of decisions remains highly correlated with the number of items on a shopping list. An e-grocery user typically purchases 21, but up to 70 unique items (Yuan et al. 2016). The average e-grocery order is 17.1 times the size of an average order on Amazon.com and 11.2 times as large as the one on Walmart.com (Yuan et al. 2016). The large order sizes are an e-grocery specific root cause for IO.

In summary, e-grocery suffers from the problem of IO. We identified user heterogeneity and large orders as two root causes. Referring to its adverse effects on the user, mitigating IO is a goal worth pursuing.

## 5. Requirement Engineering

Reflecting on the problem statement, reducing IO in e-grocery is the main goal. The application of RS is a viable approach to address IO (Aljukhadar et al. 2010; Schafer et al. 1999). Like the paper from Elsner and Krämer (2013), we build the RS design requirements subject to the domain and application specifics. We suggest one design requirement as a countermeasure for each identified root cause of IO in e-grocery. Finally, we elaborate on the recommendation technology for e-grocery as the third design requirement.

We address the root cause of large orders with a design requirement for the RS output. Since the order size is large in e-grocery and naturally varies between users and orders of the same user, we suggest providing an item ranking. This way, a user can stop scrolling through the ranking once the user found all items on the current shopping list. Any ranking should cover at least all items on a user's shopping list. Otherwise, the user is not able to complete the intended purchase. If the ranking equals the shopping list, the number of decisions is minimal. If the number of decisions is minimal, the user perceives minimal IO. We denote the length of the ranking, and the shopping list by r and s, respectively. With perfect recommendation accuracy, $r = s$ is sufficient to cover all items on the shopping list. The length of the ranking must increase with decreasing recommendation accuracy. Consequently, the shopping list length, as well as the accuracy of recommendations, determine the minimal required ranking length. Formally, the relationship is given by equation (3).

$$r \geq \frac{100\%}{\text{accuracy}} * s \qquad (3)$$

We exemplify equation (3) with the shopping list size and accuracy of the Walmart Online Grocery study by Yuan et al. (2016). In their study, the most significant order captures 70 unique items ($s = 70$). The authors measure accuracy by item hit rate. The hit rate in their study is 12%. This accuracy and shopping list length require a minimal ranking length of 584 items to cover all items on the shopping list, assuming constant accuracy. Experiments described in the literature of food RS have shown that regardless of the source of user feedback applied (i.e., ratings, tags, or comments) standard methods are only capable of producing a relatively unsatisfactory performance in preference learning (Trattner and Elsweiler 2017). The accuracy of RS in e-grocery is not only far below 100% in practice but also lower than in other RS domains. Therefore, an assumption of decreasing recommendation accuracy with the increasing length of ranking is more realistic. The drop-in accuracy further enlarges the minimal ranking length required. The length of the ranking ranges between 1 and R ($r = \{1, ..., k, ..., N, ..., R\}$). In our notation, k refers to k/R≪1%, and the following relation holds k ≪ N. Equation (3) exemplified by the Walmart Online Grocery study (Yuan et al. 2016) clarifies that top-k rankings provide a limited potential to minimize IO in e-grocery. Enlarging k to N ensures to exceed the minimal ranking length for every user and thereby the ability to reduce IO. Hence, we cover all items on a shopping list by personalized recommendations. Therefore, the first requirement is a top-N ranking.

In a shopping scenario, one ranking cannot display optimal item recommendations for two users with different preferences. No two customers' buying behavior is the same (Friedman 2011). The IO root cause of user heterogeneity calls for individual recommendations (Adomavicius et al. 2008). Consequently, the second requirement for RS in e-grocery is an individual ranking for each user. In general, a high level of individualism of recommendations - called personalization - is challenging. The major reason is data sparsity (Desrosiers and Karypis 2011). Nonetheless, in e-grocery, large and frequent orders (Yuan et al. 2016) can overcome the sparsity problem and make personalized rankings both a suitable solution and a valid requirement.

Last, a fundamental design choice concerns the recommendation technology. It influences the performance of an RS. Available technologies include collaborative, content-based, knowledge-based filtering (Aggarwal 2016). Diverse domain characteristics such as heterogeneity, preference stability, and interaction style offer different opportunities for recommendation technologies. Burke and Ramezani (2011) propose a matching strategy to select a suitable RS technology for application domains, which they describe along with six domain factors. Table 1 provides these factors and the evaluation of Burke and Ramezani (2011) for e-commerce, for which e-grocery is a subset.

Table 1 also provides the authors' evaluation of the characteristics of e-grocery, which coincides with the e-commerce results to a large extent. The only difference concerns the characteristic of churn. A high churn domain is one in which item assortment changes rapidly (Burke and Ramezani 2011). Many e-commerce domains, like fashion, face high item churn rates, while e-grocery does not. Considering all six characteristics, the high similarity between e-commerce and e-grocery implies that the most suitable RS technologies coincide with each other. Burke and Ramezani (2011) point out collaborative filtering as the best RS technology for e-commerce. Moreover, the authors highlight that collaborative filtering performs better in low churn domains. Therefore, the third requirement for an e-grocery RS is a collaborative filtering technology.

In summary, we obtain three design requirements that RS should meet to reduce IO in e-grocery: (1) top-N ranking, (2) personalized ranking, and (3) collaborative filtering.

|  | E-Grocery | E-Commerce |
|---|---|---|
| Risk | Low | Low |
| Heterogeneity | High | High |
| Preferences | Stable | Stable |
| Interaction style | Implicit | Implicit |
| Scrutability | Not required | Not required |
| Churn | Low | High |

Table 1. Domain factors of e-grocery and e-commerce

## 6. Research Gap

This section maps the state-of-the-art RS to the derived requirements. The mapping displays a research gap. Table 2 depicts how the e-grocery RS proposed in prior work perform against the requirements. The ranking length is the primary source of divergence between the requirements and existing algorithms. The existing RS produce top-k rankings with at most k=32 items instead of top-N. All nine state-of-the-art RS provide personalized rankings and thereby meet the second requirement. Most RS also use collaborative filtering and meet the third requirement. However, Table 2 confirms that no personalized top-N RS with collaborative filtering exists in the literature.

|  | Ranking Length | Personalized | Method |
| --- | --- | --- | --- |
| Requirement | Top-N | Yes | Collaborative Filtering |
| (Wan et al. 2018) | Top-32 | Yes | Collaborative Filtering |
| (Ma et al. 2019) | Top-20 | Yes | Collaborative Filtering (generic) |
| (Yuan et al. 2016) | Top-8 | Yes | Hybrid-based Filtering |
| (Sano et al. 2015) | Top-5 | Yes | Collaborative Filtering |
| (Wu and Teng 2011) | Top-3 | Yes | Collaborative Filtering |
| (Li et al. 2009) | Top-3 | Yes | Collaborative Filtering |
| (Dias et al. 2008) | Top-3 | Yes | Collaborative Filtering |
| (Panikar et al. 2016) | Top-1 | Yes | Hybrid-based Filtering |
| (El-Dosuky et al. 2012) | No information | Yes | Content-based Filtering |

Table 2.    Recommender systems for e-grocery in literature

In summary, the mapping shows that the existing RS are not meeting all requirements to address IO in e-grocery to an adequate extent. Consequently, there is a demand for a collaborative and personalized top-N RS that effectively reduces IO in e-grocery.

## 7. Operationalization of Information Overload

In e-grocery, IO depends on the number of decisions a user is taking. The number of required decisions for purchasing one item equals its rank. To purchase an item with rank 4, the user takes three negative decisions and one positive (see Figure 2). Consequently, the average number of decisions a user takes during the ordering process is the average over $\text{rank}_{i,o}$ for all added items $I_o$ by the user in order $o$. This metric is called the average rank of correct recommendation (ARC). Equation (4) gives ARC for each order $o$.

$$\text{ARC}_o = \frac{\sum_{i \in I_o} \text{rank}_{i,o}}{|I_o|} \qquad (4)$$

Averaging $\text{ARC}_o$ across all orders, $O$ results in ARC, as given in equation (5).

$$\text{ARC} = \frac{\sum_{o \in O} \text{ARC}_o}{|O|} \qquad (5)$$

Burke used ARC to capture the recommender's performance in terms of accuracy from a user's perspective (Burke 2004, 2007). The smaller the ARC, the fewer decisions are required, and the lower the information load. The difference in ARC between two algorithms is the average number of additional decisions a user faces. The minimal $\text{ARC}_o$ appears in the case of the best-case-ranking in Figure 2. Since every recommendation is correct, equation (6) gives the average rank. The dependency of $|I_o|$ demonstrates once more that the order size determines the minimal number of decisions.

$$\min(\text{ARC}_o) = \frac{|I_o| + 1}{2} \qquad (6)$$

A reduction of decisions reduces the information load of the user. Therefore, this operationalization corresponds to the definition of IO.

## 8. Design Top-N Recommender System

In this section, we build a top-N RS to cover the research demand and evaluate it offline. By definition, a design is complete and effective if it satisfies the requirements and constraints of the defined problem (Hevner et al. 2004). Applying this to our study, the RS must meet the three design requirements and reduce IO. We perform an offline evaluation based on a data set of a German e-grocer. Since this paper focuses on the online evaluation of top-k versus top-N RS to identify effects in the field, we keep the offline evaluation succinct.

### 8.1. Build E-G-COUSIN

This section builds a recommendation algorithm to reduce IO in e-grocery, which meets the deduced three requirements of being a (1) top-N ranking, (2) personalized ranking, and based on (3) collaborative filtering. First, we select an algorithm fulfilling these requirements from the literature. Secondly, the algorithm is adapted to e-grocery to reduce IO further.

The algorithm selection starts with the most restrictive requirement, the collaborative recommendation technology. We further reduce the solution space of collaborative filtering methods to those that provide personalized top-N rankings. Thus, we exclude popular collaborative filtering methods like association rule mining or probabilistic approaches such as Bayes Classifier. Three collaborative filtering methods can calculate personalized top-N rankings: matrix-factorization, user-based, and item-based collaborative filtering. Gan (2015) proposes a collaborative filtering algorithm, called COUSIN, that combines user-based and item-based filtering, and includes user, item, and user-item associations simultaneously. These features capitalize on the substantial order information in e-grocery. The developers of COUSIN also demonstrate its superiority over other collaborative filtering methods in an offline evaluation (Gan 2015). Based on the requirement fit and the evaluation results, we consider COUSIN to be an appropriate candidate RS in e-grocery and as a subject to domain-specific modifications.

COUSIN is designed to generate movie recommendations. In contrast, e-grocery has different characteristics, and the algorithm design should address these. One necessary transfer from the movie to the e-grocery domain is the operationalization of user preferences. Gan (2015) considers the existence of a positive user rating to be a statement of preferences (e.g., if user $u_1$ rates movie $i_1$ positive). In e-grocery, item rating is not holistically established. Several e-grocers such as Walmart Online Grocery (Walmart.com 2021) have no user rating at all. Besides, purchases are a suitable and always available implicit indication of preference (Ahrens 2011; Neumann and Geyer-Schulz 2008).

Therefore, we equate user preferences in e-grocery with item purchases. Furthermore, we propose two algorithmic adaptions of COUSIN to increase recommendation accuracy for e-grocery: (1) user-item-matrix split and (2) no power-law-adjustment. The adaptions change the input to the algorithm (adaption (1)) and the data transformation within the algorithm (adaption (2)). Note that these modifications do not change the fundamental functionalities of COUSIN. The abbreviation of COUSIN's e-grocery adaption is E-G-COUSIN.

Figure 3 is a schematic visualization of E-G-COUSIN for which step (3) till step (6) corresponds to the original workflow of COUSIN (Gan 2015). Step (1) and (2) highlight the adaptions in black.

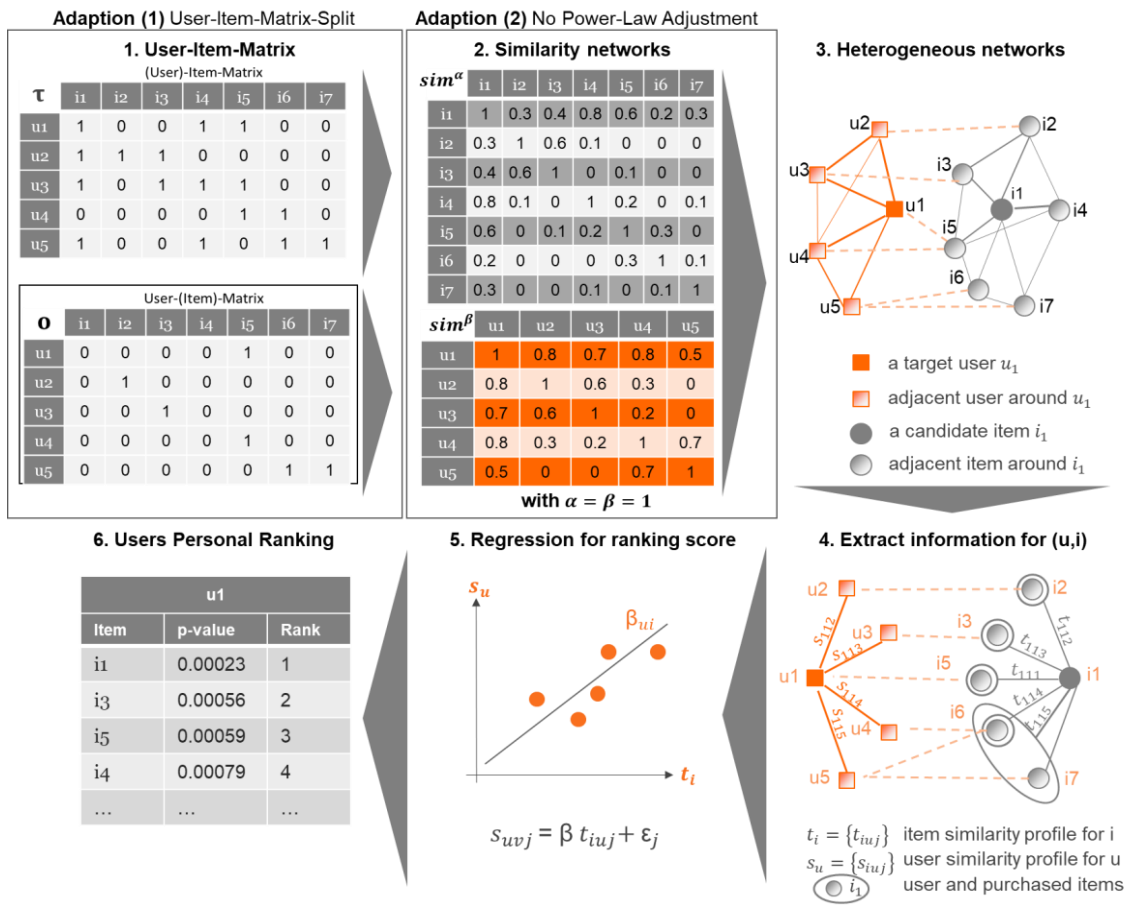

Figure 3.    E-G-COUSIN - Adaption of COUSIN (Gan 2015) for e-grocery

E-G-COUSIN embodies six phases: (1) extract user-item-matrices based on the meta-parameters $\tau$ and $o$; (2) calculate cosine similarity between items and between users; (3) construct a heterogeneous network out of user similarities, item similarities, and user-item transactions from the user-(item)-matrix; (4) construct a subnetwork for all user-item combinations (u, i); (5)

perform regression through the origin for each (u, i); (6) rank items for each user by the p-value of the regression f-test. We refrain from a detailed description of COUSIN, which is available in (Gan 2015). Rather, we elaborate on the two adaptions that we propose for e-grocery.

**Adaption 1:** User-Item-Matrix Split

We suggest two separate user-item-matrices to calculate user and item similarities. The objective of the user-item-matrix split is to provide a high accuracy of recommendations and high coverage of users. We hypothesize COUSIN to suffer from the general RS trade-off between accuracy and user coverage (Aggarwal 2016) in e-grocery. The reason for the trade-off is the single user-item-matrix used for both similarity calculations. COUSIN can only provide one set of transactions in the user-item-matrix, e.g., recent transactions or an extended period. The quality of item similarities benefits from recent user-item transactions, due to seasonal and trend effects in food. Therefore, a user-item-matrix with recent transactions improves recommendation accuracy, but only the full transaction data set can provide 100% user coverage.

Step (1) in Figure 3 shows the user-item-matrix split of E-G-COUSIN. Each user-item-matrix has its meta-parameter that determines the transaction data inclusion. These tuneable meta-parameters identify the ideal information base for item and user similarities. The (user)-item-matrix employs meta-parameter $\tau$. Thus, the (user)-item-matrix contains transaction data of the past $\tau$ days to calculate item similarity. The user-(item)-matrix employs meta-parameter $o \in (0,100]$, which specifies the empirical quantile of orders in percent regulating the maximal amount of orders included in the user-(item)-matrix for each user. Due to $o>0$, the minimum amount of orders per user is 1.

**Adaption 2:** No Power-Law-Adjustment

Gan (2015) suggests a power-law adjustment of similarities. The aim is to magnify the difference between strong and weak similarities effectively. Gan (2015) assumes strong similarities resulting from sharing a significant fraction of preferred items, weak similarities resulting from sharing a small number of preferred items. Based on this assumption, a power-law-adjustment is an advisable transformation. The similarity (sim) transformation is performed by $sim^\alpha$, with α>1 for item similarities and $sim^\beta$, with β>1 for user similarities. So far, this assumption is not applicable to e-grocery. Strong similarities are likely to result by sharing top-seller items that are visible to a large user base, due to the low level of recommendation individuality. The maximum level of individuality in research is a top-32 ranking (Wan et al. 2018). As a result, large shares of users receive the same item recommendations by a default ranking. Generally, users are more likely to buy an item from the default ranking they see, due to the "default effect". This cognitive bias says individuals tend to remain with their default choice even in case the default is randomly assigned (Kessler and Zhang 2017). This effect leads to high purchasing frequencies of the top items in default ranking. Consequently, buying these generates strong similarities based on factitious preferences. Power-law-adjustment would strengthen these strong similarities, such that the default items are ranked top again. This way, the new ranking approximates the default one, referred to as overfitting. We aim to avoid overfitting of the ranking by not performing a power-law adjustment. Thus, we do not transform the similarities by setting $\alpha = \beta = 1$.

### 8.2. Evaluate E-G-COUSIN

By design, E-G-COUSIN fulfills all three design requirements, i.e., top-N ranking, personalized ranking, and collaborative filtering. To verify the design complete- and effectiveness of E-G-COUSIN, we performed an offline evaluation by comparing E-G-COUSIN and COUSIN on a real-world data set from a German e-grocer. The 28 June 2017 serves as an evaluation day with 1,003 users and 21,654 transactions. The offline evaluation confirmed that E-G-COUSIN reduces information load in terms of ARC aggregated over all users by 7.9 %. These results prove both adaptions to be effective in practice and justifies the further application of E-G-COUSIN within our study.

## 9. Field Testing

Field testing involves the real-life implementation of a software package (Tomei 2007). The top-k RS (E-G-NN) and top-N RS (E-G-COUSIN) are the two software packages we study here. Real-life implementation at the e-grocer means that the sorting of the full range supermarket assortment (~22,000 items) in the shop corresponds to the ranking of each RS.

In the following, we specify the evaluation design at a German e-grocer. Additionally, we present a collaborative and personalized top-k RS to benchmark IO in the field. Afterwards, we quantify the effects of top-N versus top-k RS on IO by the real-world data obtained. Finally, we summarize additional user and company effects that have also been provoked by the collaborative and personalized top-N RS in the field.

### 9.1. Evaluation Design

An evaluation design consists of three characteristics: a subject, setting, and research method (Jannach et al. 2011). The subject of our evaluation design was the repeat and active e-grocery user. The restriction to repeat users is in the nature of any collaborative filtering algorithm due to the cold-start problem (Jannach et al. 2011). Please note that solving the cold-start problem has been out of scope for this paper. We define users as active if they have purchased within one year before the test start. The evaluation setting was a field study. A field study means that we run the experiment in a real-world scenario. The cooperation with the e-grocer allowed us to conduct an online evaluation as an experimental research method. Online evaluation can assess the comparative performance of various algorithms (Aggarwal 2016), in this case, top-k vs top-N. No standardized or specialized online evaluation protocols exist for food RS (Trattner and Elsweiler 2017). One popular application of online evaluation on webpages is A/B testing (Sauro and Lewis 2016). A/B tests measure the direct impact of the RS on the user and are the most accurate approach for testing the long-term performance of the RS (Aggarwal 2016). We performed an A/B test for evaluating the effects of top-N versus top-k RS in the field.

The e-grocer provided access to users according to the definition of evaluation subjects. For the A/B test, we randomly assigned the users into two groups. Group A received the top-k ranking and group B the top-N ranking. A user belongs to the same group during the test. The test period was 17 August till 19 September 2017. All other conditions (e.g., voucher logic and newsletter content) stayed the same for users of both groups during the test. Table 3 presents the number of users per group.

| Group | Ranking | Users |
|-------|---------|-------|
| A | top-k | 12,218 |
| B | top-N | 12,309 |

Table 3.    A/B test – Group assignment

We require a ranking of all ~22,000 items for both RS to provide access to the full assortment. In each full assortment ranking, the k or N personalized items were displayed first. The remaining items were ranked based on the same cluster popularity ranking. E-G-NN provided personalized top-k rankings with k=7 on average. E-G-COUSIN provided personalized top-4000 rankings. The rankings applied to all pages presenting lists with more than 16 items in the shop. These pages included all category pages, subcategory pages, and topic-specific landing pages like sales. Consequently, the ranking on these pages was a subset of the full assortment ranking. For example, if a user entered the sales section, the user received all sale items ranked according to the RS ranking. The search results and carousels with top-16 recommendations were not affected by the ranking.

In summary, we performed an A/B test at the e-grocer for one month to evaluate E-G-COUSIN with top-4000 against E-G-NN with top-7 on average. In total, 24,527 repeat and active e-grocery users participate in the test.

## 9.2. Benchmark Top-k Recommender System

We use a collaborative, personalized, top-k ranking as a state-of-the-art benchmark since it is the most frequent state-of-the-art RS design in Table 2. E-grocery users purchase items repeatedly since groceries are fast-moving consumer goods. Against this background, items purchased before are highly relevant to a user in the future. An RS with recently and repeatedly purchased items provide personal top-k recommendations. This method is a user-based nearest-neighbor collaborative filtering for which the nearest neighbor is the user itself. We abbreviate this approach with E-G-NN. E-G-NN provides a top-k ranking with k=7 on average, if k≥1, which is true for 89.1% of the customers. The value of k varies based on user purchasing history. A frequent customer with a rich purchasing history might have k=30. This top-k ranking represents a state-of-the-art RS in e-grocery because it meets the three requirements top-k ranking, personalized ranking, and collaborative filtering, like the majority of current RS, proposed for e-grocery in Table 2. E-G-NN is an example of a real-world top-k RS since the e-grocer used it in their webshop in 2016 and 2017.

### 9.3. Effect on Information Overload

In the following, we test if a shift from a collaborative and personalized top-k to our top-N RS reduces IO in e-grocery. This evaluation allows validating the top-N ranking requirement in the field. Moreover, we extend the knowledge base by measuring the real-world effect of a top-N RS on IO reduction in e-grocery.

The A/B test at the e-grocer with top-k RS E-G-NN and top-N RS E-G-COUSIN provides the necessary field test data. We test if top-N reduced IO in terms of ARC with the hypothesis $H_0$ and $H_1$ given by equations (7) and (8).

$$H_0: \text{ARC}_{top-k} \leq \text{ARC}_{top-N} \tag{7}$$

$$H_1: \text{ARC}_{top-k} > \text{ARC}_{top-N} \tag{8}$$

$\text{ARC}_{top-k}$ averages the number of decisions a user takes by receiving the top-k ranking. $\text{ARC}_{top-N}$ quantifies the average number of decisions of a user shopping with a top-N ranking. In the case of $H_1: \text{ARC}_{top-k} > \text{ARC}_{top-N}$ the top-N ranking requires fewer decisions and thereby provides a lower information load.

Table 4 summarizes the empirical A/B test results for ARC during the online evaluation. One order is one observation. We used the item rank in the ranking of the complete assortment to calculate ARC. In this way, we assess the performance of the RS independent of effects by navigation behavior.

| Group | Ranking | Orders | ARC (mean) | Standard deviation of ARC | p-value |
|---|---|---|---|---|---|
| A | top-k | 563 | 3454.4 | 2639.0 | <0.01 |
| B | top-N | 608 | 2437.9 | 2084.0 | |
| | | | | | Data: 30 August 2017 |

Table 4.     A/B test – Effects on information overload

Our field test showed that group A users took 3454.4 decisions on average to purchase one item, while group B users required 1016.5 decisions less. These results show that the top-N ranking reduced IO by 29.4% in the real-world scenario. We applied a two-sample Welch's t-test to test the hypothesis in equations (7) and (8). Welch test has been applicable because the assumption of a normal distribution of ARC in both groups is valid (see Appendix 1). This statistical test evaluates the hypothesis while accounting for variance heterogeneity (Quinn and Keough 2002). We reject $H_0: \text{ARC}_{top-k} \leq \text{ARC}_{top-N}$, based on the p-value in Table 4. Consequently, we consider the IO reduction based on the shift from the collaborative and personalized top-k to our top-N ranking as highly significant.

### 9.4. Effect on Customer Perceived and Business Value

Additions to the knowledge base as results of design science research also include all experiences gained from performing the research and field testing in the application environment (Hevner 2007). The following section addresses this demand, in particular, because evaluations of food RS in literature have mostly been offline with proprietary collections (Trattner and Elsweiler 2017). Against this background, we provide additional findings on customer perceived and business value by E-G-COUSIN obtained in the field testing.

CPV increased in the dimension of usability, also called ease-of-use (Gummerus 2010; Jiang et al. 2016). The usability improved due to search time reduction. We use the time between two add-to-basket actions as an indicator of search time. The longer the time between two add-to-basket actions, the longer the user runs through the ranking before finding the next item to buy. The median time in seconds between two add-to-basket actions reduced significantly by 3.3 seconds and the mean by 10 seconds.

| Group | Ranking | Conversion in terms of orders per user | | Seconds between two add-to-basket actions | | | Number of purchased distinct items | | |
|---|---|---|---|---|---|---|---|---|---|
| | | Users | Orders | Mean | Median | STD | Mean | Median | STD |
| A | top-k | 12,218 | 10026 | 152.2 | 67 | 908.6 | 19.95 | 17 | 14.8 |
| B | top-N | 12,309 | 10648 | 142.2 | 63.7 | 591.1 | 20.33 | 18 | 15.1 |
| Statistical test: p-value | | Chi-squared independence: <0.01 | | Mood's median test: <0.01 | | | Two-sample t-test: <0.05 | | |
| | | | | | | | Data:17 August 2017 till 19 September 2017 | | |

Table 5.    A/B test –Effects on users and company

Moreover, users' usability increased by finding additional relevant items. The number of purchased distinct items per order in group B significantly increased compared to group A. Precisely group B users bought one item more in median and 0.38 in the mean. This sales increase indicates that a user finds more items on the shopping list through the top-N ranking than through the top-k ranking. Table 5 provides the mean, median, and standard deviation (STD) for both CPV measures.

The company benefited from a significant increase in revenue. The revenue development bases on the increase in order volume and value. In group A 12,218 users made 10,026 orders. Group B made 6.2% more orders while being only 0.7% more users. This increase results in a significant conversion lift of 5.4%. Another effect for the e-grocer was that the average order value of group B is 1.36 € higher than for group A. Both effects aggregate to a 7% revenue increase. This lift in customer perceived and business value demonstrates that our collaborative and personalized top-N RS E-G-COUSIN has a significant positive practical impact besides IO reduction.

## 10. Discussion

This paper demonstrates that a shift from collaborative and personalized top-k to top-N RS can significantly reduce IO in e-grocery. In this way, we provided the first field study on the effect of short (top-k) and long (top-N) rankings in e-grocery. In the following, we discuss the results and embed our findings in related research. Next, we describe limitations and suggest future research.

By now, e-grocery RS in the literature focus on top-k rankings. Our research carries their work forward and embeds the present knowledge about e-grocery characteristics to derive specific RS design requirements. Based on these requirements, we contributed the first top-N RS for e-grocery and successfully evaluated it in the field. This finding enables e-grocers to implement an algorithm that already once reduced IO and increased revenue significantly in the field. The existing literature on e-grocery RS uses classification metrics to evaluate their top-k RS in the field. In that, they neither measure the overall information load nor evaluate the effect size of their RS on IO in e-grocery. Our e-grocery study addressed this gap.

We provided evidence that a personalized and collaborative top-N RS can reduce IO significantly compared to top-k. The online evaluation at an e-grocer delivered high reliability of our findings. Moreover, our online evaluation is a major advantage since research widely lacks online studies to evaluate the real-world effects of RS. Only a few online evaluations of e-grocery RS exist in the literature. Those studies never gave insights on algorithmic details, IO effects as well as user and company benefit at the same time. Providing this comprehensive picture opens a new door for research. Not only RS research, but the whole information systems community can leverage the results to design more effective artifacts and formulate novel hypotheses for behavioral studies. Moreover, the delivered evidence about user and company benefits motivates prospective integration of top-N RS in the e-grocery context.

Academic RS research generally aims to reduce IO and primarily uses accuracy metrics (Aggarwal 2016). However, the literature lacks a direct link between information load and metrics. In this paper, we built the link between the rank accuracy measure ARC and the information load for the first time. We identified that the amount of processed information is proportional to the number of decisions a user takes. Therefore, the information load equals the rank position. This reading provides excellent interpretability in absolute and relative values. Absolute values correspond to the average number of decisions a user must take to purchase one item. The relative value serves as a comparison of two levels of information load. This interpretation has been an unnoticed advantage of ARC and demonstrates its function to operationalize IO.

The primary limitation of this study is generalizability. Concrete, the identified effects are limited to the tested algorithms and e-grocery company. Our study missed the evaluation at a variance of e-grocery companies to ensure generality. Nevertheless, the e-grocer in our field test represented a classic supermarket assortment with over 20,000 items, large orders with an average of 22 distinct items, and high user heterogeneity. Therefore, we assume that our results hold for e-grocers with similar characteristics. A second aspect is a limited generalizability on personalized and collaborative top-k and top-N RS. The evaluation covered one algorithm for each ranking length, which limits the robustness of the results. However, both algorithms were representative of their kind. The top-k RS E-G-NN was a real-world example and best-practice at the e-grocer. A similar top-N RS like E-G-COUSIN has demonstrated superiority over state-of-the-art recommendation techniques in the movie domain (Gan 2015). Therefore, it is likely that future research can verify our findings in additional e-grocery studies.

Moreover, our top-N RS for e-grocery, E-G-COUSIN, has two major limitations. First, E-G-COUSIN has not minimized the information load. We stopped the design process when we met the requirements. The average of 2437.9 decisions per item indicates that there is still a high potential for IO reduction. Secondly, E-G-COUSIN has not addressed the cold-start-problem. Therefore, new users face the same IO as before.

We suggest three main areas for future research. First, research should provide additional studies in e-grocery to generalize our findings. Moreover, these evaluations can provide an estimation of the functional relationship between ranking length and information load. Secondly, behavioral studies should investigate the effects of top-k and top-N RS as well as the role of IO in the field. Thirdly, we invite the research community to develop and improve top-N RS for e-grocery, since we have demonstrated their enormous potential.

## 11. Conclusion

IO is a problem in grocery shopping with negative effects on customers and business performance. User heterogeneity and large orders are two significant characteristics that aggravate IO in a grocery context. The rise of e-grocery provides the opportunity to reduce IO by applying RS to grocery shopping. To that end IO in e-grocery, we derived three design requirements for RS - top-N ranking, personalized ranking, and collaborative filtering. A mapping of state-of-the-art RS against these requirements identified a research gap. We addressed this gap by designing the first personalized and collaborative top-N RS for e-grocery. We evaluated our RS with N=4000 offline and online in the field with a randomized experiment at a German e-grocer. The benchmark supplied a personalized and collaborative top-k RS with k=~7. Both rankings, top-7 and top-4000, were complemented by a semi-personalized-ranking of all remaining items in the shop. The A/B test demonstrated a significant reduction of IO through the top-N RS. The field test also revealed significant value lifts for both, customers and the business from applying our top-N RS.

This paper achieved the objective of IO reduction in e-grocery. Moreover, this study provides one contribution for each design science research cycle. Contributions are summarized in Figure 4.

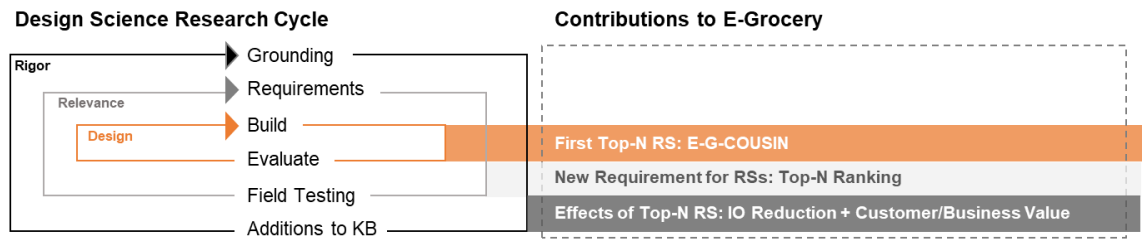

Figure 4.   Design science research cycles and contributions

First, we provided an offline and online evaluated top-N RS for e-grocery, called E-G-COUSIN. Secondly, we submitted a new design requirement for e-grocery RS, i.e., top-N ranking. Thirdly, we quantified various effects of the top-N RS in the field. The top-N RS significantly reduced IO in comparison to a top-k RS by 29.4%. Moreover, we demonstrated that our top-N RS increases business value by 7% revenue in the field, as well as customer value by a significant reduction of user search time by 3.3 seconds per item. With these contributions, we raise the awareness of the high potential of top-N RS for e-grocery. Furthermore, corporate practice does now have the chance to implement E-G-COUSIN to lift significant value for customers and business.

# 12. References

Abu ELSamen AA (2015) Online Service Quality and Brand Equity: The Mediational Roles of Perceived Value and Customer Satisfaction. Journal of Internet Commerce 14:509–530. https://doi.org/10.1080/15332861.2015.1109987

Adomavicius G, Huang Z, Tuzhilin A (2008) Personalization and Recommender Systems. In: Chen Z-L, Raghavan S (eds) Tutorials in operations research: State-of-the-art decision-making tools in the information-intensive age. presented at the INFORMS Annual Meeting, October 12-15, 2008. INFORMS, Hanover, pp 55–107

Aggarwal CC (2016) Recommender systems: The textbook. Springer, Cham

Ahrens SC (2011) Recommender systems: Relevance in the consumer purchasing process. Epubli, Berlin

Aljukhadar M, Senecal S, Daoust C-E (2010) Information Overload and Usage of Recommendations. In: Bart P. Knijnenburg, Lars Schmidt-Thieme, Dirk Bollen (ed) Proceedings of the ACM 2010 Workshop on User-Centric Evaluation of Recommender Systems and Their Interfaces, vol 5. ACM, pp 26–33

Burke R (2004) Hybrid Recommender Systems with Case-Based Components. In: Funk P, González Calero PA (eds) Advances in case-based reasoning, vol 3155. Springer, Berlin, London, pp 91–105

Burke R (2007) Hybrid Web Recommender Systems. In: Brusilovsky P, Kobsa A, Nejdl W (eds) The adaptive web: Methods and strategies of Web personalization, vol 4321. Springer, Berlin, pp 377–408

Burke R, Ramezani M (2011) Matching Recommendation Technologies and Domains. In: Ricci F, Rokach L, Shapira B, Kantor PB (eds) Recommender systems handbook. Springer, New York, pp 367–386

Catalina (2013) Engaging the Selective Shopper: Why Today's Consumers Expect Personalization. https://assets.contentful.com/oyvkhyupj8l5/9bcwE3FVQW6gMWm8O8QgG/7c87c2ea1cb5c368fc2215bd563d06ad/SSfinal_Study.pdf. Accessed 3 January 2021

Chang H-H, Wang H-W (2011) The moderating effect of customer perceived value on online shopping behaviour. Online Information Review 35:333–359. https://doi.org/10.1108/14684521111151414

Chen Z, Dubinsky AJ (2003) A conceptual model of perceived customer value in e-commerce: A preliminary investigation. Psychol. Mark. 20:323–347. https://doi.org/10.1002/mar.10076

Delen D (2015) Real-world data mining: Applied business analytics and decision making. Pearson, Upper Saddle River, NJ

Delen D, Ram S (2018) Research challenges and opportunities in business analytics. Journal of Business Analytics 1:2–12. https://doi.org/10.1080/2573234X.2018.1507324

Desrosiers C, Karypis G (2011) A Comprehensive Survey of Neighborhood-based Recommendation Methods. In: Ricci F, Rokach L, Shapira B, Kantor PB (eds) Recommender systems handbook. Springer, New York, pp 107–144

Dias MB, Locher D, Li M, El-Deredy W, Lisboa PJG (2008) The Value of Personalised Recommender Systems to e-Business: A Case Study. In: Pu P, Bridge D, Mobasher B, Ricci F (eds) Proceedings of the ACM 2008 Conference on Recommender Systems. ACM, New York, pp 291–294

El-Dosuky MA, Rashad MZ, Hamza TT, EL-Bassiouny AH (2012) Food Recommendation Using Ontology and Heuristics. In: Hassanien AE, Salem A-BM, Ramadan R, Kim T-h (eds) Advanced Machine Learning Technologies and Applications, vol 322. Springer Berlin Heidelberg, Berlin, Heidelberg, pp 423–429

Elsner H, Krämer J (2013) Managing Corporate Portal Usage with Recommender Systems. BISE (Business & Information Systems Engineering) 5:213–225. https://doi.org/10.1007/s12599-013-0275-3

Eppler MJ, Mengis J (2004) The Concept of Information Overload: A Review of Literature from Organization Science, Accounting, Marketing, MIS, and Related Disciplines. The Information Society 20:325–344. https://doi.org/10.1080/01972240490507974

Friedman HJ (2011) No thanks, I'm just looking: Sales techniques for turning shoppers into buyers. Wiley, Hoboken, N.J

Gan M (2015) COUSIN: A network-based regression model for personalized recommendations. Decision Support Systems 82:58–68. https://doi.org/10.1016/j.dss.2015.12.001

Gummerus J (2010) E-services as resources in customer value creation: A service logic approach. Managing Service Quality 20:425–439. https://doi.org/10.1108/09604521011073722

Hevner A, March S, Park J, Ram S, Hevner, March, Park, Ram (2004) Design Science in Information Systems Research. MIS Quarterly 28:75–105. https://doi.org/10.2307/25148625

Hevner AR (2007) A Three Cycle View of Design Science Research. Scandinavian Journal of Information Systems 19:87–92

Hevner AR, Chatterjee S (2010) Design research in information systems: Theory and practice. Integrated series in information systems, vol 22. Springer, New York, NY

Jacoby J, Speller DE, Kohn CA (1974a) Brand Choice Behavior as a Function of Information Load. Journal of Marketing Research 11:63. https://doi.org/10.2307/3150994

Jacoby J, Speller DE, Berning CK (1974b) Brand Choice Behavior as a Function of Information Load: Replication and Extension. Journal of Consumer Research 1:33–42. https://doi.org/10.1086/208579

Jannach D, Zanker M, Felfernig A, Friedrich G (2011) Recommender systems: An Introduction. Cambridge University Press, Cambridge

Jannach D, Zanker M, Ge M, Gröning M (2012) Recommender Systems in Computer Science and Information Systems – A Landscape of Research. In: Huemer C, Lops P (eds) E-Commerce and Web Technologies, vol 123. Springer, Heidelberg, pp 76–87

Jiang L, Jun M, Yang Z (2016) Customer-perceived value and loyalty: How do key service quality dimensions matter in the context of B2C e-commerce? Serv Bus 10:301–317. https://doi.org/10.1007/s11628-015-0269-y

Kessler JB, Zhang YC (2017) Behavioural economics and health. In: Detels R, Gulliford M, Karim QA, Tan CC (eds) Oxford Textbook of Global Public Health, 6th edn. Oxford University Press, Oxford, pp 775–789

Kuo Y-F, Wu C-M, Deng W-J (2009) The relationships among service quality, perceived value, customer satisfaction, and post-purchase intention in mobile value-added services. Computers in Human Behavior 25:887–896. https://doi.org/10.1016/j.chb.2009.03.003

Lawrence RD, Almasi GS, Kotlyar V, Viveros MS, Duri SS (2001) Personalization of Supermarket Product Recommendations. Data Mining and Knowledge Discovery 5:11–32. https://doi.org/10.1023/A:1009835726774

Levitin DJ (2015) The Organized Mind: Thinking straight in the age of information overload. Penguin UK, London

Li M, Dias BM, Jarman I, El-Deredy W, Lisboa PJG (2009) Grocery shopping recommendations based on basket-sensitive random walk. In: Elder J (ed) Proceedings of the 15th ACM SIGKDD International Conference on Knowledge Discovery and Data Mining. ACM, New York, pp 1215–1224

Lien C-H, Wen M-J, Huang L-C, Wu K-L (2015) Online hotel booking: The effects of brand image, price, trust and value on purchase intentions. Asia Pacific Management Review 20:210–218. https://doi.org/10.1016/j.apmrv.2015.03.005

Ma L, Cho JHD, Kumar S, Achan K (2019) Seasonality-Adjusted Conceptual-Relevancy-Aware Recommender System in Online Groceries. In: Baru C (ed) IEEE International Conference on Big Data. IEEE, Piscataway, NJ, pp 4435-4443

Melinat P, Kreuzkam T, Stamer D (2014) Information Overload: A Systematic Literature Review. In: Johansson B, Andersson B, Holmberg N (eds) Perspectives in Business Informatics Research, vol 194. Springer, Cham, pp 72–86

Milord JT, Perry RP (1977) A Methodological Study of Overload. J Gen Psychol 97:131–137. https://doi.org/10.1080/00221309.1977.9918509

Neumann AW, Geyer-Schulz A (2008) Applying Small Sample Test Statistics for Behavior-based Recommendations. In: Preisach C, Burkhardt H, Schmidt-Thieme L, Decker R (eds) Data analysis, machine learning and applications. Springer, Berlin, pp 541–549

Panikar S, Mane P, Pakhale C, Fulzele S, Rathi A (2016) Online Grocery Recommendation System. International Journal of Scientific Research and Development 4:318–321

Patnaik LM, Hiriyannaiah S (2017) Business Analytics Using Recommendation Systems. In: Mandal JK, Dutta P, Mukhopadhyay S (eds), vol 775. Springer, Singapore, pp 35–44

PitchBook (2017) Online grocery shopping sales in the United States from 2012 to 2021 (in billion U.S. dollars). https://www.statista.com/statistics/754815/online-grocery-consumer-share/. Accessed 15 January 2020

Quinn G, Keough M (2002) Experimental design and data analysis for biologists. Cambridge University Press, Cambridge

Sano N, Machino N, Yada K, Suzuki T (2015) Recommendation System for Grocery Store Considering Data Sparsity. Procedia Computer Science 60:1406–1413. https://doi.org/10.1016/j.procs.2015.08.216

Sauro J, Lewis JR (2016) Quantifying the user experience: Practical statistics for user research. Elsevier, Waltham, MA

Schafer B, Konstan J, Riedl J (1999) Recommender Systems in E-Commerce. In: Feldman S (ed) Proceedings of the 1st ACM conference on Electronic commerce. ACM, New York, pp 158–166

Tomei LA (2007) Online and Distance Learning: Concepts, methodologies, tools and applications. IGI Global, Hershey, PA

Walmart.com (2021) Walmart Online Grocery. https://www.walmart.com/grocery/. Accessed 3 January 2021

Wan M, Wang D, Liu J, Bennett P, McAuley J (2018) Representing and Recommending Shopping Baskets with Complementarity, Compatibility and Loyalty. In: Cuzzocrea A, Schuster A, Wang H, Allan J, Paton N, Srivastava D, Agrawal R, Broder A, Zaki M, Candan S, Labrinidis A (eds) Proceedings of the 27th ACM International Conference on Information and Knowledge Management. ACM Press, New York, USA, pp 1133–1142

Wu Y-J, Teng W-G (2011) An enhanced recommendation scheme for online grocery shopping. In: The 15th IEEE International Symposium on Consumer Electronics. Institute of Electrical and Electronics Engineers; IEEE, Piscataway, NJ, pp 410–415

Yang Z, Peterson RT (2004) Customer perceived value, satisfaction, and loyalty: The role of switching costs. Psychol. Mark. 21:799–822. https://doi.org/10.1002/mar.20030

Yuan M, Pavlidis Y, Jain M, Caster K (2016) Walmart Online Grocery Personalization: Behavioral Insights and Basket Recommendations. In: Link S, Trujillo J (eds) Advances in conceptual modeling: ER 2016 Workshops, AHA, MoBiD, MORE-BI, MReBA, QMMQ, SCME, and WM2SP, vol 9975. Springer, Cham, pp 49–64

# Appendix

**Appendix 1:** Normal Distribution Assumption for Average Rank of Correct Recommendation per Test Group

**Average Rank of Correct Recommendation (ARC)**

ARC – Group A 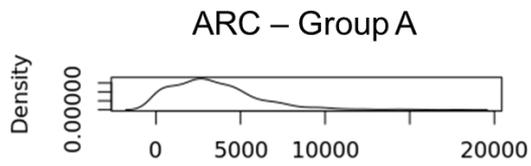

ARC – Group B 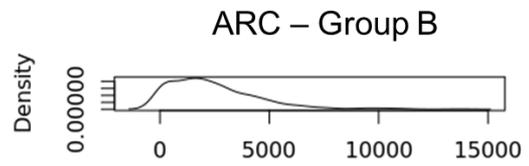